# The Study of Hydrophilicity and Optical Properties of Zn and N-doped TiO$_2$ Thin Films

**Manish K Vishwakarma**[*,1], **Manjeet**[1], **and P Jain**[1]

[1]Department of Physics, Indian Institute of Technology Roorkee, Roorkee, Uttarakhand, India, 247667

*Corresponding author: mvishwakarma@ph.iitr.ac.in*

**Abstract**

Among the wide bandgap semiconductors, TiO$_2$ is the highly stable and cost-efficient semiconductor used for the different photocatalysis processes like water splitting, chemical waste degradation, anti-micro bacterial activities, and more. Materials showing high hydrophilicity (surface phenomenon) with low bandgap are required to improve photocatalysis efficiency. We report the synthesis of Zn (4 wt.%) and N (4 wt.%) doped TiO$_2$ thin films using the spin coating technique to improve surface wettability. The XRD pattern shows the growth of the pure anatase phase of TiO$_2$. UV absorption spectra show a minor increment in the bandgap of the Zn and N doped TiO$_2$ thin films. The water contact angle with pure TiO$_2$ is 33.45° and reduces to 17.94° after 4wt% doping of Zn and N. The results show the enhanced hydrophilicity in the Zn and N doped TiO$_2$ thin films.

## Introduction

The hydrophilicity of the surface is an essential property of a photocatalyst. Metal oxides (TiO$_2$[1], ZnO, NiO, and more) were always popular among researchers in wide bandgap semiconductors. TiO$_2$ is a chemically stable, cost-efficient metal oxide that shows high photocatalytic property[2]. In this paper, we proclaim the synthesis of Zn and N-doped TiO$_2$ using the spin coating technique. Our focus is to study the change in the surface wettability and optical properties of the TiO$_2$ after the transition metal (Zn) and non-metal (N) doping. The water contact angle measurement is performed to check the surfaces' hydrophilic property, and UV-absorption spectroscopy is done to study the change in optical properties. Crystallinity and compactness of the surface are studied with the help of XRD measurement and FE-SEM images, respectively.

## Thin Film Fabrication

Pristine and doped TiO$_2$ thin films are prepared using the spin coating technique. Precursor solutions were prepared for the p-TiO$_2$ thin films by dissolving 1 ml of titanium tetra-isopropoxide (TTIP) in 5 ml of isopropyl alcohol (IPA) and stirred on a magnetic stirrer for 5 min. Then after 0.5 ml of acetic acid is mixed with the solution dropwise to prevent the precipitation of the TiO$_2$ and stirred vigorously for 1 hour. A solution is prepared for the Zn and N doping by dissolving 4 wt.% (CH3COO)$_2$Zn in methanol and 4 wt.% Urea in IPA, respectively. The prepared dopant solution is mixed accordingly with the precursor and coated on the glass substrate (1 cm x 1 cm x 1.34 cm) with the help of the spin coater at 3000 rpm for 60 sec. Coated films were dried on the hot plate at 100 for 1H and then annealed in a muffle furnace at 450°$C$ in air for 1H to achieve the crystallinity and remove the organic solvents.

## Results and Discussion

Figure 1 represents the XRD pattern of p-TiO$_2$ and Zn, N-doped TiO$_2$. The Pure anatase phase of TiO$_2$ is formed with the two peaks at $2\theta = 25.3°$ and $2\theta = 47.9°$. The crystallite size is calculated using Scherrer's equation,

$$D = \frac{0.9\lambda}{\beta Cos\theta}$$

Where D is the crystallite size, $\lambda$ is the wavelength of the X-Ray, $\beta$ is the FWHM of the peak, and $\theta$ is the angle. We found that the crystallite size got decreased with the doping. Crystallite size of p-TiO$_2$ and Zn, N-TiO$_2$ at peak positions $2\theta = 25.3°$ are 0.2558 nm, 0.2957 nm.

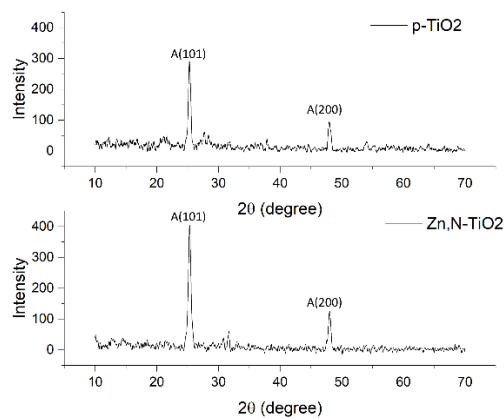

*Figure 1 XRD pattern of p-TiO$_2$ and Zn, N-doped TiO$_2$*

Figure 2 represents the FE-SEM micrographs of p-TiO$_2$ (Figure 2(a)) and Zn, N-TiO$_2$ (Figure 2(b)) which shows the flat and compact surface morphology. The lattice distortion and the change in the grain size of the crystallite are not obvious from the micrographs.

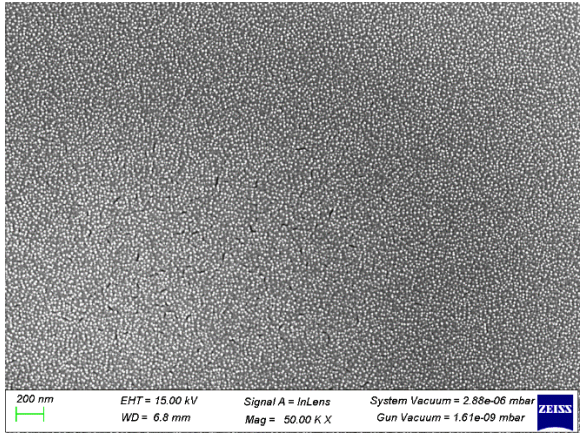

*Figure 2a p-TiO$_2$*

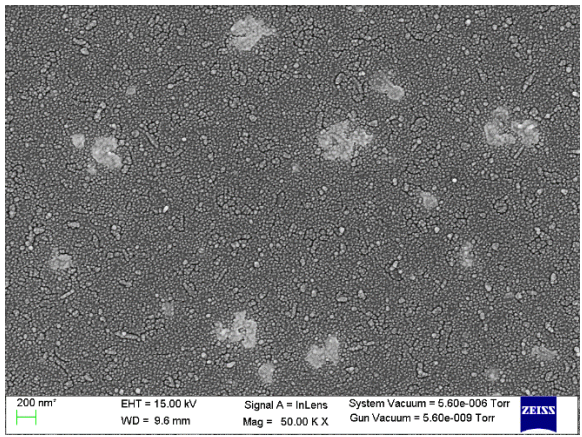

*Figure 3(b) Zn, N-TiO$_2$*

*Figure 3 FESEM micrographs of TiO$_2$ Thin Films*

Figure 3 (a) shows the UV visible absorption spectra of p-TiO$_2$ and Zn, N-TiO$_2$ thin films. We calculated the optical bandgap of the p-TiO2 and Zn, N-TiO$_2$ thin films 3.19 eV and 3.48 eV, respectively using the tauc plot method. The absorption spectra of Zn, N-TiO$_2$ show the blue shift with respect to p-TiO$_2$. Figure 4 shows the water contact angle (WCA) with the Zn, N-TiO$_2$ thin film. WCA with Zn, N-TiO$_2$, and p-TiO$_2$ are 31.45° and 17.94°, respectively, which shows the doping of Zn and N increased the surface hydrophilicity impressively.

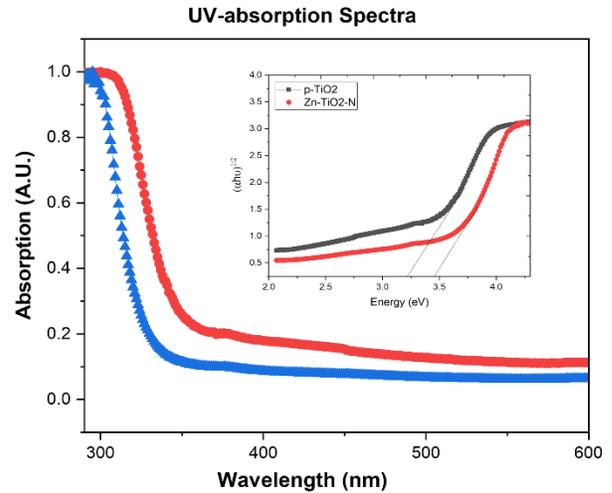

*Figure 3 UV Vis absorption spectra and tauc plot*

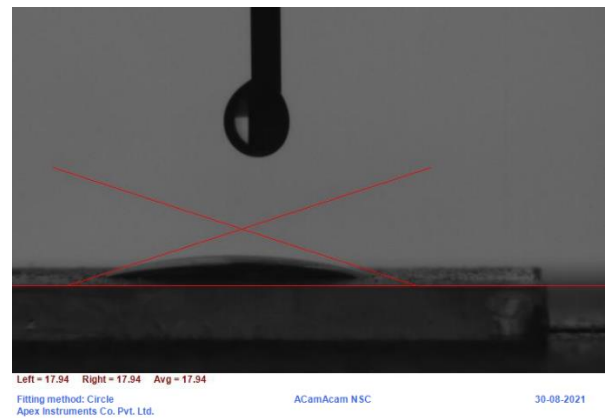

*Figure 4 WCA of Zn, N-TiO$_2$*

## Conclusion and Acknowledgement

In this paper, we synthesized the undoped TiO$_2$ and Zn (4 wt.%) and N (4 wt.%) doped TiO$_2$ thin films and characterized it systematically. We found an impressive increment in hydrophilicity. We also reported the observed blue shift in the absorption spectra of the Zn, N-TiO$_2$. Finally, I would like to thank my colleague Mr. Ramesh Kumar for his support and discussions regarding the work.

## References

1. M Tang Bo-Wen, Niu Shan et.al. Ferroelectrics 2019, Vol. 549, 96–103.

2. Abdullah M. Alotaibi and Benjamin A. D. Williamson, et. al. ACS Appl. Mater. Interfaces 2020, 12, 13, 15348–15361.